\documentclass[aps,twocolumn]{revtex4}
\usepackage{graphicx}  
\usepackage{dcolumn}   
\usepackage{bm}        
\usepackage{amssymb}   
\usepackage{hyperref}
\usepackage{tikz}
\usetikzlibrary{shapes,arrows}
\usepackage{amsmath}

\begin{document}

\title{Unruh effect using Doppler shift method in DSR framework}
\author{Vishnu Rajagopal}
\email{vishnurajagopal.anayath@gmail.com}
\affiliation{Deenbandhu Chhotu Ram University of Science and Technology\\Murthal, Sonepat, Haryana - 131039, India}

\begin{abstract}

We study the Unruh effect in doubly special relativity (DSR) framework by generalising the Doppler-shift method to DSR. For both the scalar and Dirac particles, we observe a deviation in the power spectrum of Unruh radiation from the standard Bose-Einstein and Fermi-Dirac distributions, respectively, due to the presence of the frame independent length scale of DSR. We further show that this deviation results in the modification of Unruh temperature which then depends non-linearly on the proper acceleration in DSR. 

\end{abstract}

\maketitle

\section{Introduction}

The existence of the minimal length scale of the order of Planck scale is a crucial ingredient of the quantum gravity studies \cite{ml}. The concept of such a frame independent length scale necessitates the possibility of extending the special theory of relativity to the doubly special relativity (DSR) \cite{dsr}. Unlike the special theory of relativity, the DSR possess two frame independent quantities: the velocity of light and the minimal length scale \cite{cam1}. A major consequence of this length scale is in the modification of the energy-momentum dispersion relation and as a result this modified dispersion relation contain terms that depend on the Planck mass \cite{mdr,smolin}. The minimal length scale also results in the break down of the space-time continuum and leads to the emergence of non-commutative (NC) space-time structure \cite{snyder,doplicher}.  

Various types of NC space-times such as Moyal space-time \cite{moyal}, $\kappa$-Minkowski space-time \cite{kappa-review}, etc., have been thoroughly explored in the recent times. Of these, the $\kappa$-Minkowski space-time (with a Lie algebraic type of non-commutativity) has been shown to be the underlying space-time of the DSR whose symmetry algebra is defined in terms of the $\kappa$-Poincare algebra (which is a deformed Poincare algebra) \cite{kpa,kpa1}. The $\kappa$-Poincare algebra contains an additional observer independent scale $\kappa$ (which is related to Planck mass) such that in the limit $\kappa\to\infty$ the $\kappa$-Poincare algebra can be contracted to obtain the standard Poincare algebra. Thus a consistent DSR theory can be constructed rigorously by exploiting the $\kappa$-Poincare algebra, (which has a Hopf algebraic structure) and the modified dispersion relation associated with the DSR can be obtained from the Casimir of the $\kappa$-Poincare algebra \cite{kpa2}. Thus the $\kappa$-Poincare algebra in different basis will give arise to different modified dispersion relations (which are the deformations of the standard dispersion relation). But it is to be noted that all these different modified dispersion relations are equivalent as they are obtained from the representations of the same $\kappa$-Poincare algebra in different basis, which are inter related through non-linear transformations \cite{nowak}. At this point it is worth mentioning that the symmetry algebra of the $\kappa$-Minkowski space-time can also be realised using an undeformed $\kappa$-Poincare algebra which contain an undeformed Poincare algebra and a deformed co-algebra sectors \cite{meljanac}.  

The studies related with the black hole and the associated Hawking radiation \cite{hawking}, which is the thermal radiation of black hole observed by an inertial observer in the vicinity of event horizon, can a provide plausible insight of the quantum gravity effects. The phenomenon of Hawking radiation is closely analogues to the Unruh effect \cite{unruh} where a uniformly accelerating observer will find the vacuum to be in a thermal bath which radiates with a temperature (known as Unruh temperature) linearly related to the uniform acceleration. Thus the studies on Unruh effect within the context of NC space-times will also shed some light into quantum gravity signals. 

In the recent times, various studies have been initiated to understand the effects of non-commutativity on Unruh effect. The response function of a uniformly accelerating monopole detector coupled to the scalar \cite{kim,ravi1} and Dirac \cite{ravi2} fields, satisfying $\kappa$-deformed equations of motion, has been shown to deviate from the standard thermal distribution of Unruh temperature. The $\kappa$-deformation on the Unruh effect has been analysed in \cite{vr2} by calculating the expectation value of the number operator for scalar field in Minkowski vacuum by employing the $\kappa$-deformed oscillator algebra in the method of Bogoliubov coefficients. The $\kappa$-deformed corrections to the power spectrum of the outgoing scalar modes detected by a Rindler observer has been obtained in \cite{glikman} using the star product formalism compatible with $\kappa$-Poincare algebra and in \cite{vr1} using the plane wave mode solution of the $\kappa$-Klein-Gordon equation compatible with the undeformed $\kappa$-Poincare algebra. In the recent times various other works along these direction have been studied within the $\kappa$-Minkowski space-time \cite{referee}.   

In the present work, we compute the $\kappa$-deformed corrections of Unruh effect, for both scalar and Dirac particles separately, by generalising the Doppler shift method discussed in \cite{paul} to DSR. Here we choose to calculate the Doppler shifted frequency in DSR using the $\kappa$-Lorentz transformation associated with the $\kappa$-NC phase space compatible with the $\kappa$-Poincare algebra in the Magueijo–Smolin (MS) base \cite{nowak}. Using this Doppler shifted frequency one can obtain the expression for the plane wave mode moving along the uniformly accelerating trajectory \cite{paddy1}. The Fourier transform of this plane wave mode is the major ingredient for calculating the power spectrum detected by a uniformly accelerating observer and this power spectrum $\mathcal{P}(\Omega)$ is related to the Fourier mode $\mathcal{F}(\Omega)$ as $\mathcal{P}(\Omega)=\Omega|\mathcal{F}(\Omega)|^2$ (see \cite{paddy1} for details). In this work, the notion of power spectrum plays a crucial role in studying Unruh effect in DSR as this gives the DSR corrections of the standard thermal distribution functions associated with the Unruh effect.

This paper is organised in the following manner. In Sec. II, we provide a brief discussion on the $\kappa$-NC phase algebra and the $\kappa$-Lorentz transformation associated with the DSR. We also discuss the Darboux mapping (and its inverse) that connects the $\kappa$-NC and the standard commutative phase variables. In Sec. III, we study the DSR Doppler effect and the associated red shift by deriving the expression for the Doppler shifted frequency in DSR using the $\kappa$-Lorentz transformation. In Sec. IV, we use this DSR Doppler shifted and obtain the expression of plane wave mode for the scalar and Dirac particles separately. In Subsec. A and Subsec. B, we calculate the thermal power spectrum of Unruh effect for scalar and Dirac particles in the DSR framework respectively. Finally in Sec. V, we discuss our results and provide the concluding remarks.

\section{$\kappa$-Lorentz transformation}

In this work we study the Unruh effect using the Doppler shift approach within the DSR framework and therefore our analysis is purely classical for the time being. For this reason we choose the DSR framework discussed in \cite{subir1,subir2,subir3,subir4} which provides a classical description using the Poisson brackets among the $\kappa$ phase space variables. Thus the $\kappa$-Minkowski non-commutative (NC) phase space algebra in the covariant form is given as
\begin{equation}\label{a}
\begin{split}
 \{x_{\mu},x_{\nu}\}=&\frac{1}{\kappa}(x_{\mu}\eta_{\nu 0}-x_{\nu}\eta_{\mu 0}),~~\{p_{\mu},p_{\nu}\}=0,\\
 \{x_{\mu},p_{\nu}\}=&-\eta_{\mu\nu}+\frac{1}{\kappa}\eta_{\mu 0}p_{\nu},
\end{split}
\end{equation}
where the signature of $\eta_{\mu\nu}$ is $(+,-,-,-)$. We recover the canonical phase space algebra when the kappa deformation energy scale blows to infinity, i.e., $\kappa\to\infty$.

The Lorentz generator for this DSR framework is defined as \cite{subir2}
\begin{equation}\label{b}
 J_{\mu\nu}=x_{\mu}p_{\nu}-x_{\nu}p_{\mu}.
\end{equation}
Using the above definition (i.e., Eq.(\ref{b})) and the $\kappa$-NC algebra given in Eq.(\ref{a}), we observe that Lorentz algebra remains intact under the $\kappa$ deformation , i.e.,
\begin{equation}\label{c}
\{J_{\mu\nu},J_{\alpha\beta}\}=\eta_{\mu\beta}J_{\nu\alpha}+\eta_{\mu\alpha}J_{\beta\nu}+\eta_{\nu\beta}J_{\alpha\mu}+\eta_{\nu\alpha}J_{\mu\beta}.
\end{equation}
Under the action of the above Lorentz generator, the $\kappa$-NC phase space coordinates $x_{\mu}$ and $p_{\mu}$ transform as
\begin{equation}\label{d}
\begin{split}
 \{J_{\mu\nu},x_{\rho}\}&=\eta_{\nu\rho}x_{\mu}-\eta_{\mu\rho}x_{\nu}+\frac{1}{\kappa}(p_{\mu}\eta_{\nu 0}-p_{\nu}\eta_{\mu 0})x_{\rho},\\
 \{J_{\mu\nu},p_{\rho}\}&=\eta_{\nu\rho}p_{\mu}-\eta_{\mu\rho}p_{\nu}+\frac{1}{\kappa}(p_{\mu}\eta_{\nu 0}-p_{\nu}\eta_{\mu 0})p_{\rho}.
\end{split}
\end{equation} 
One can see that the $\delta x_{\mu}$ and $\delta p_{\nu}$ (which are the infinitesimal changes in $x_{\mu}$ and $p_{\mu}$) associated with the transformation in Eq.(\ref{d}) are modified due to the $\kappa$-NC phase algebra defined in Eq.(\ref{a}).

Now using Eq.(\ref{a}),Eq.(\ref{c}) and Eq.(\ref{d}), we write down the explicit form of the $\kappa$-Poincare algebra (in the MS basis) \cite{subir2} as
\begin{equation}\label{d0}
\begin{split}
 \{J_{\mu\nu},J_{\alpha\beta}\}&=\eta_{\mu\beta}J_{\nu\alpha}+\eta_{\mu\alpha}J_{\beta\nu}+\eta_{\nu\beta}J_{\alpha\mu}+\eta_{\nu\alpha}J_{\mu\beta},\\
 \{J_{\mu\nu},p_{\rho}\}&=\eta_{\nu\rho}p_{\mu}-\eta_{\mu\rho}p_{\nu}+\frac{1}{\kappa}(p_{\mu}\eta_{\nu 0}-p_{\nu}\eta_{\mu 0})p_{\rho},\\
 \{p_{\mu},p_{\nu}\}&=0.
\end{split}
\end{equation}
From the above arguments and Eq.(\ref{d0}) it is clear that the Lorentz transformation for the $\kappa$-NC phase space coordinates are modified due to the underlying $\kappa$-Poincare algebra. This modified Lorentz transformation compatible with the $\kappa$-Poincare algebra is known as the $\kappa$-Lorentz transformation \cite{subir2}. Hence under the action of $\kappa$-Lorentz transformation, the momenta vector $(\varepsilon,p_x,p_y,p_z)$ transform as (see \cite{subir2} for the complete derivation)
\begin{equation}\label{d1}
\begin{split}
 \varepsilon'&=\frac{\gamma(\varepsilon-vp_x)}{1+\frac{1}{\kappa}((\gamma-1)\varepsilon-\gamma vp_x)},\\
 p_x'&=\frac{\gamma(p_x-v\varepsilon)}{1+\frac{1}{\kappa}((\gamma-1)\varepsilon-\gamma vp_x)},\\
 p_y'&=\frac{p_y}{1+\frac{1}{\kappa}((\gamma-1)\varepsilon-\gamma vp_x)},\\
 p_z'&=\frac{p_z}{1+\frac{1}{\kappa}((\gamma-1)\varepsilon-\gamma vp_x)}.
\end{split}
\end{equation}
From the above it is clear that $\kappa$-Lorentz transformation reduces to the usual Lorentz transformation in the limit $\kappa\to\infty$.

Therefore it is intuitive to think about a mapping from the usual Lorentz transformation to the $\kappa$-Lorentz transformation and vice-versa. For this purpose we introduce the canonical phase space variables $(X_{\mu},P_{\mu})$ satisfying
\begin{equation}\label{d2}
\begin{split}
 \{X_{\mu},X_{\nu}\}=0,~~\{P_{\mu},P_{\nu}\}=0,~~\{X_{\mu},P_{\nu}\}=&-\eta_{\mu\nu}.
\end{split}
\end{equation}
The $\kappa$-NC phase space variables $(x_{\mu},p_{\mu})$ (having non-canonical Poisson bracket) can be mapped to the canonical phase space variables $(X_{\mu},P_{\mu})$ using the following transformation law (Darboux map) \cite{subir2,subir3,subir4}
\begin{equation}\label{d3}
\begin{split}
 x_{\mu}=X_{\mu}\Big(1+\frac{E}{\kappa}\Big),~~p_{\mu}=\frac{P_{\mu}}{\Big(1+\frac{E}{\kappa}\Big)}.
\end{split}
\end{equation}
Similarly the corresponding inverse transformation law (inverse Darboux map) is given as \cite{subir2,subir3,subir4}
\begin{equation}\label{e}
\begin{split}
 X_{\mu}=x_{\mu}\Big(1-\frac{\varepsilon}{\kappa}\Big),~~P_{\mu}=\frac{p_{\mu}}{\Big(1-\frac{\varepsilon}{\kappa}\Big)}.
\end{split}
\end{equation}
By substituting Eq.(\ref{d3}) in Eq.(\ref{d1}), we obtain the usual Lorentz transformation law for the canonical momenta vector $(E,P_x,P_y,P_z)$ as
\begin{equation}\label{d4}
\begin{split}
 E'&={\gamma(E-vP_x)},~~P_x'={\gamma(P_x-vE)},\\
 P_y'&=P_y,~~P_z'=P_z.
\end{split}
\end{equation}
Thus by using the above defined inverse mapping (i.e., Eq.(\ref{e})) in the usual Lorentz transformation, we obtain the $\kappa$-Lorentz transformation for the momenta vector, as shown in Eq.(\ref{d1}).

In many cases it is quite difficult to construct the unique Lagrangian, compatible with the $\kappa$-NC phase algebra (given in Eq.(\ref{a})), for a particle in the DSR framework. Under these conditions it is straightforward to construct the Lagrangian using the usual canonical phase space variables and then utilising the inverse map given in Eq.(\ref{e}), we will get the Lagrangian in terms of the $\kappa$-NC variables (see \cite{subir3,subir4} for details). We will be using this approach in the upcoming sections to study Unruh effect in the DSR framework.


\section{Doppler effect in DSR framework}

Here we study the doubly special relativistic Doppler effect by calculating the general expression for the Doppler shifted frequency compatible with the DSR using $\kappa$-Lorentz transformation. We then obtain an expression for the time-dependent Doppler effect by calculating the Doppler shifted frequency, detected by a uniformly accelerating observer and we do this in two ways. In the first method, we obtain this by employing the inverse mapping of the momenta coordinate (defined in Eq.(\ref{e})) in the usual Lorentz transformation and in the second method, we derive the same expression by directly using $\kappa$-Lorentz transformation. 

The doubly special relativistic Doppler effect for the frequency (a massless plane wave) can be obtained directly from the expression for the $\kappa$-Lorentz transformation of the momenta vector (given in Eq.(\ref{d1})) as
\begin{equation}\label{e3}
 \omega'=\frac{\omega\gamma(1-v)}{1-\frac{\omega}{\kappa}(1-\gamma(1-v))}\equiv\frac{\omega\sqrt{\frac{1-v}{1+v}}}{1-\frac{\omega}{\kappa}\Big(1-\sqrt{\frac{1-v}{1+v}}\Big)}.
\end{equation}
The above equation is the general expression for the doubly special relativistic Doppler shifted frequency and this will reduce to the familiar relativistic Doppler shifted frequency in the limit $\kappa\to\infty$. Note that in the above $v$ is the relative velocity between the observer and source, where $v>1$ when the observer is moving away from the source and $v<1$ when the observer is moving towards the source. 

Another interesting quantity that can be obtained from the relativistic Doppler effect is the red-shift \cite{weinberg} and it is defined as $z\equiv\frac{\Delta\lambda}{\lambda}=\frac{\lambda'-\lambda}{\lambda}$ \cite{weinberg}, where $\lambda'$ is the wavelength detected by the observer in a frame moving with velocity $v$ and $\lambda$ is the wavelength detected by the observer in rest frame. In terms of the frequency this can be re-written as $z=\frac{\omega-\omega'}{\omega'}$. Thus by substituting Eq.(\ref{e3}) in the definition of the redshift, we obtain the expression for the DSR redshift as 
\begin{equation}\label{red}
 1+z=\sqrt{\frac{1+v}{1-v}}\bigg(1-\frac{\omega}{\kappa}\Big(1-\sqrt{\frac{1-v}{1+v}}\Big)\bigg).
\end{equation}
From the above expression (i.e., Eq.(\ref{red})), we observe that the redshift in DSR depend on the frequency of the emitted photon $\omega$ as well as on the energy scale $\kappa$, unlike the standard redshift (caused due to the special relativistic Doppler effect) which depends only on the relative velocity between the source and the observer. In Fig.(\ref{fig:unrh}) we show the variation of $1+z$ against the relative velocity. The dotted line (having $\omega/\kappa=0.01$) corresponds to the case of DSR and solid line (having $\omega/\kappa=0$) corresponds to the case of special relativity.  

\begin{figure}[!htb]\centering
\includegraphics[width=0.48\textwidth]{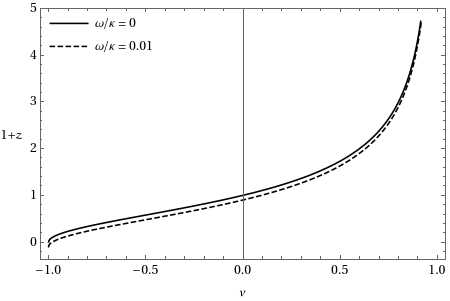}
\caption{Variation of redshift against relative velocity}
\label{fig:unrh}
\end{figure}

We begin our analysis, to obtain an expression for the time dependent Doppler frequency, by considering a uniformly accelerating observer (Rindler observer) and the trajectory of this Rindler observer is defined in the following manner
\begin{equation}\label{e1}
 T(\tau)=\frac{1}{A}\sinh A\tau,~~X(\tau)=\frac{1}{A}\cosh A\tau,
\end{equation}
where $A$ is the constant proper acceleration and $\tau$ is the proper time. Similarly the velocity of the accelerating observer detected by the Minkowski observer is defined as $v=\frac{dX}{dT}$. Using Eq.(\ref{e1}) in $v=\frac{dX}{dT}=\frac{dX}{d\tau}\frac{d\tau}{dT}$, we get the velocity as $v(\tau)=\tanh A\tau$.

Now let us consider a plane wave of frequency $W$ and wave vector $\vec{K}$ (moving along the $X$ direction). In the instantaneous rest frame of the observer moving with velocity $v$, the frequency $W'$ of the plane wave detected by this observer is given using the Lorentz transformation corresponding to momenta vector given in Eq.(\ref{d4}). Thus we have
\begin{equation}\label{e2}
 W'=\gamma(W-vK_x).
\end{equation}
The frequency $W'$ of the massless plane wave (for which $W=|\vec{K}|$) detected by an uniformly accelerating observer (whose trajectory is Eq.(\ref{e1})), can be calculated from Eq.(\ref{e2}) by setting the velocity of the frame to be $v=\tanh A\tau$ and this gives 
\begin{equation}\label{e5}
 W'=We^{-A\tau}.
\end{equation}
In the above expression, $W'$ can be interpreted as the Doppler shifted frequency detected by the uniformly accelerating observer and from the RHS we find that this time-dependent Doppler frequency decreases exponentially with time \cite{paul}.

We now apply the inverse mapping (defined for the momenta vector, i.e., Eq.(\ref{e})) in Eq.(\ref{e5}). Thus its LHS and RHS become $W'=\frac{\omega'}{1-\omega'/\kappa}$ and $We^{-A\tau}=\frac{\omega}{1-\omega/\kappa}e^{-A\tau}$ respectively. By equating these terms and re-arranging them, we get the frequency $\omega'$ detected by the Rindler observer in DSR framework as 
\begin{equation}\label{e4}
 \omega'=\frac{\omega e^{-A\tau}}{1-\frac{\omega}{\kappa}(1-e^{-A\tau})}.
\end{equation}
From the above expression, we notice that in the limit $\kappa\to\infty$, the $\omega'$ (in Eq.(\ref{e4})) reduces to the $W'$ (given in Eq.(\ref{e5})), as expected. Under the small acceleration limit i.e., $A\tau<<1$, the Doppler shifted frequency in DSR can be approximated as $\omega'\simeq\frac{\omega(1-A\tau)}{1-\omega A\tau/\kappa}$. 

We can also re-derive Eq.(\ref{e4}) directly from the $\kappa$-Lorentz transformation. For this we need to substitute $v(\tau)=\tanh A\tau$ in Eq.(\ref{e3}) and this will give the expression for the DSR Doppler shifted frequency, detected by the Rindler observer.

\section{Unruh effect from Doppler shifted frequency in DSR}

In this section we study the Unruh effect in DSR by employing the Doppler shift method discussed in \cite{paul}. For this we first obtain the expression for an outgoing wave moving with a DSR Doppler frequency (given in Eq.(\ref{e4})) and then calculate the corresponding power spectrum detected by the Rindler observer, using the Fourier transform of the plane wave mode. We will calculate this power spectrum in DSR separately for the scalar and Dirac particles respectively.

The phase factor $\phi(\tau)$ of the plane wave corresponding to this doubly special relativistic Doppler shifted frequency can be obtained using the definition $\phi(\tau)=\int\omega'(\tau')d\tau'$ \cite{paul}. Substituting Eq.(\ref{e4}) in this definition and performing the integration, we get $\phi(\tau)$ as
\begin{equation}\label{f1}
 \phi(\tau)=\frac{-\kappa}{A}\ln\Big(1+\frac{\omega}{\kappa}(e^{-A\tau}-1)\Big)+C,
\end{equation}
where $C$ is the constant of integration. Similarly the phase factor for the standard special relativistic Doppler shifted frequency $W'$ (given in Eq.(\ref{e5})) is found to be $\Phi(\tau)=-\frac{W}{A}e^{-A\tau}$ \cite{paul}. Now we fix $C$ in Eq.(\ref{f1}) by comparing $\Phi(\tau)=-\frac{W}{A}e^{-A\tau}$ with the commutative limit of Eq.(\ref{f1}), i.e., $\lim_{\kappa\to\infty}\phi(\tau)$. It is to be noted that $W\to\omega$ in commutative limit where $\kappa\to\infty$. By using the identity $\lim_{y\to\infty}\ln(1+\frac{1}{y})^y=1$ and then setting $W\to\omega$ (for the limit $\kappa\to\infty$), Eq.(\ref{f1}) becomes $\lim_{\kappa\to\infty}\phi(\tau)=\phi_{(\kappa\to\infty)}(\tau)=-\frac{W}{A}(e^{-A\tau}-1)+C$. Thus by comparing $\Phi(\tau)$ with $\phi_{(\kappa\to\infty)}(\tau)$, we fix the value of $C$ to be $C=-\frac{W}{A}$. But we need to transform this $W$ back to $\omega$ before substituting the value of $C$ in Eq.(\ref{f1}) and this can be be done using the inverse mapping given in Eq.(\ref{e}) which gives $W=\frac{\omega}{1-\omega/\kappa}$. Thus we get $C=-\frac{\omega}{A(1-\omega/\kappa)}$ and substituting this in Eq.(\ref{f1}), we get the phase factor in DSR as 
\begin{equation}\label{f2}
 \phi(\tau)=\frac{-\kappa}{A}\ln\Big(1+\frac{\omega}{\kappa}(e^{-A\tau}-1)\Big)-\frac{\omega}{A(1-\frac{\omega}{\kappa})}.
\end{equation}
We now use this phase factor (compatible with DSR) and obtain the plane wave expression for the scalar and Dirac particles respectively. We will further use it to study the Unruh effect for bosons and fermions in DSR.  

\subsection{Scalar particle}

The wave $\psi(\tau)$ corresponding to a phase factor $\phi(\tau)$ can be written as $\psi(\tau)=e^{i\phi(\tau)}$. Thus using the phase given in Eq.(\ref{f2}), we get the scalar wave associated with the DSR Doppler shifted frequency (i.e., Eq.(\ref{e4})) as
\begin{equation}\label{f3}
 \psi(\tau)=\exp\bigg(-\frac{i\kappa}{A}\ln\Big(1+\frac{\omega}{\kappa}(e^{-A\tau}-1)\Big)-\frac{i\omega}{A(1-\frac{\omega}{\kappa})}\bigg).
\end{equation}
We need the Fourier transform of the mode to calculate its power spectrum. So we first write down the Fourier transformation of the mode  using the definition $\mathcal{F}(\Omega)=\int\psi(\tau)e^{i\Omega\tau}d\tau$ and substituting Eq.(\ref{f3}) in this, we get
\begin{equation}\label{f4}
 \mathcal{F}(\Omega)=\int_{-\infty}^{\infty}\exp\bigg(i\Omega\tau-\frac{i\kappa}{A}\ln\Big(1+\frac{\omega}{\kappa}(e^{-A\tau}-1)\Big)-\frac{i\omega}{A(1-\frac{\omega}{\kappa})}\bigg)d\tau.
\end{equation}
In order to evaluate the above integral, we first use the substitution $y=e^{-A\tau}$ and then by using the result $\int_0^{\infty}x^{\mu-1}(1+Cx)^{-\nu}dx=C^{-\mu}\beta(\mu,\nu-\mu)$ \cite{table}, one obtains the Fourier transform of the mode as 
\begin{equation}\label{f6}
 \mathcal{F}(\Omega)=\frac{1}{A}e^{-\frac{i\omega}{A(1-\frac{\omega}{\kappa})}}\Big(1-\frac{\omega}{\kappa}\Big)^{-\frac{i\kappa}{A}-\frac{i\Omega}{A}}\Big(\frac{\omega}{\kappa}\Big)^{\frac{i\Omega}{A}}\beta\Big(-\frac{i\Omega}{A},\frac{i\kappa}{A}+\frac{i\Omega}{A}\Big).
\end{equation}
By taking the modulus square of this Fourier transform and by using the definition of the power spectrum, i.e., $\mathcal{P}(\Omega)=\Omega|\mathcal{F}(\Omega)|^2$ \cite{paddy1}, we obtain the expression for power spectrum in DSR corresponding to the scalar mode (given in Eq.(\ref{f3})) as
\begin{equation}\label{f7}
\begin{split}
\mathcal{P}_{\kappa}(\Omega)=&\frac{2\pi}{A}\frac{1}{e^{2\pi\Omega/A}-1}\times\\&\bigg[\frac{1-e^{-2\pi\kappa/A}}{\big(1+\Omega/\kappa\big)\big(1-e^{-2\pi\kappa/A}e^{-2\pi\Omega/A}\big)}\bigg].
\end{split}
\end{equation}
The power spectrum obtained in Eq.(\ref{f7}), for the outgoing scalar modes, detected by the Rindler observer in the DSR framework is found to deviate from the familiar Bose-Einstein distribution factor, i.e., $(e^{2\pi\Omega/A}-1)^{-1}$ due to the $\kappa$ dependent terms obtained in the square bracket. It is to be noted that these factors inside the square bracket will reduce to unity when we approach the special relativistic limit (i.e., $\kappa\to\infty$) and we will recover the Bose-Einstein distribution factor as expression for the standard power spectrum.

\begin{figure}[!htb]\centering
\includegraphics[width=0.50\textwidth]{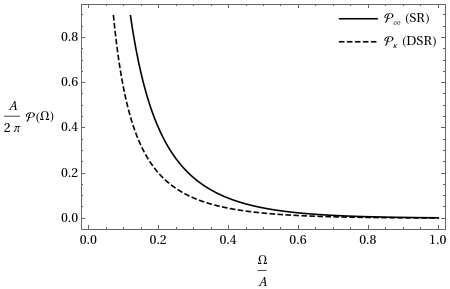}
\caption{Plot of the power spectrum for the outgoing scalar mode in DSR detected by a Rindler observer} 
\label{fig:unrh}
\end{figure}  

Fig.(\ref{fig:unrh}) shows the variation of $\frac{A}{2\pi}\mathcal{P}_{\kappa}(\Omega)$ against $\frac{\Omega}{A}$. The dotted line represents the case of DSR (where we set $\frac{A}{\kappa}\sim 10^{-3}$) and the solid line depicts the special relativistic case (where $\kappa\to\infty$). Note that we denote the power spectrum in the special relativity as $\mathcal{P}_{\infty}$. In this figure we observe a noticeable deviation of the power spectrum (given in Eq.(\ref{f7}) from the standard Bose-Einstein distribution, particularly in the small frequency limit and in the asymptotic limit, the dotted line is seen to coincide with the solid ones, as expected. We also find that the effect of the $\kappa$-energy scale (of the DSR) is to reduce the value of the power spectrum when compared with its counterpart in the special relativity.  

In general, the temperature of the thermal bath associated with the power spectrum having either a Bose-Einstein distribution (i.e., $\mathcal{P}(\Omega)=\frac{2\pi}{A}(e^{\Omega/T}-1)^{-1}$) or a Fermi-Dirac distribution (i.e., $\mathcal{P}(\Omega)=\frac{2\pi}{A}(e^{\Omega/T}+1)^{-1}$) can be read-off from this power spectrum as
\begin{equation}\label{f8}
 T=\frac{\Omega}{\ln\Big(\frac{2\pi}{A}\frac{1}{\mathcal{P}(\Omega)}\pm 1\Big)},
\end{equation}
where `$+$' and `$-$' sign in Eq.(\ref{f8}) correspond to the case of Bose-Einstein and Fermi-Dirac distributions, respectively. In Fig.(\ref{fig:unrh}) we have seen that the deviation of the modified power spectrum, i.e., Eq.(\ref{f7}) from the standard Bose-Einstein distribution becomes smaller in the larger frequency limit. Therefore it is reasonable to assume $\mathcal{P}_{\kappa}(\Omega)$ to have a thermal behaviour in the large frequency limit and hence we can read-off the corresponding temperature associated with this power spectrum using Eq.(\ref{f8}). Note that the above technique has been utilised to calculate the modified Unruh temperature associated with non-local theories \cite{gim}. At this stage it is worth mentioning that similar assumptions have been incorporated in \cite{scard} to obtain the  GUP (generalised uncertainty principle) corrections for the Unruh temperature. We will be using the above discussed procedure to calculate the modified temperature in the DSR scenario. 

In order to obtain the temperature associated with the power spectrum for the outgoing scalar modes, detected by the Rindler observer in the DSR framework, we substitute Eq.(\ref{f7}) in Eq.(\ref{f8}) and on simplifying this, we get
\begin{equation}\label{f9}
\begin{split}
 T_{\kappa}=&\frac{A}{2\pi}+\\
 &\frac{A^2}{4\pi^2\Omega}\ln\Bigg[1+\frac{\Big(\big(1+\Omega/\kappa\big)e^{-2\pi\Omega/A}-1\Big)e^{-2\pi\kappa/A}-\Omega/\kappa}{\big(1+\Omega/\kappa\big)\big(1-e^{-2\pi\Omega/A}e^{-2\pi\kappa/A}\big)}\Bigg].
\end{split}
\end{equation} 
In the above expression (i.e., Eq.(\ref{f9})), $\frac{A}{2\pi}$ represents the standard Unruh temperature and $T_{\kappa}$ represents the modified temperature of the Bose-Einstein distribution in the DSR framework. Unlike the usual Unruh temperature, here the $T_{\kappa}$ depends upon on the energy of the system. The $\kappa$-dependent correction terms of the $T_{\kappa}$ has a non-linear dependence on the proper acceleration. Note that in the limit $\kappa\to\infty$, the terms inside the square bracket of Eq.(\ref{f9}) reduce to unity and thus one obtains the usual expression for the Unruh temperature.

\subsection{Dirac particle}

Unlike the scalar particle, here we need to do the Fermi-Walker transport of the Dirac field along the uniformly accelerating trajectory \cite{kip}. Thus the Fermi-Walker transported Dirac field is given as $\tilde{\psi}(\tau)=S(\tau)\psi(\tau)$, where $S(\tau)$ is the transformation matrix responsible for this Fermi-Walker transport . For a uniformly accelerating trajectory, this transformation matrix is given as $S(\tau)=e^{-A\tau/2}$ \cite{paul}. In the $\kappa$-deformed non-commutative space-time, this transformation matrix takes the same form as that in the commutative space-time \cite{ravi2}. Thus the plane wave for the Dirac particle with the DSR Doppler shifted frequency can be obtained by substituting Eq.(\ref{f3}) and using $S(\tau)=e^{-A\tau/2}$ in $\tilde{\psi}(\tau)=S(\tau)\psi(\tau)$. Hence we get $\tilde{\psi}(\tau)$ as
\begin{equation}\label{g1}
 \tilde{\psi}(\tau)=\exp\Big(-\frac{A\tau}{2}-\frac{i\kappa}{A}\ln\Big(1+\frac{\omega}{\kappa}(e^{-A\tau}-1)\Big)-\frac{i\omega}{A(1-\frac{\omega}{\kappa})}\bigg).   
\end{equation} 
As in the case of scalar mode, now we calculate the Fourier transform associated with this mode using the definition $\mathcal{F}(\Omega)=\int \tilde{\psi}(\tau)e^{i\Omega\tau}d\tau$. By following the previous steps, as we did after Eq.(\ref{f4}), we get the Fourier transform, corresponding to the Dirac modes with DSR Doppler shifted frequency, as
\begin{equation}\label{g2}
\begin{split}
 \mathcal{F}(\Omega)=&\frac{1}{A}e^{-\frac{i\omega}{A(1-\frac{\omega}{\kappa})}}\Big(1-\frac{\omega}{\kappa}\Big)^{-\frac{i\kappa}{A}-\frac{i\Omega}{A}+\frac{1}{2}}\Big(\frac{\omega}{\kappa}\Big)^{-\frac{i\Omega}{A}-\frac{1}{2}}\times\\
 &\beta\Big(-\frac{i\Omega}{A}+\frac{1}{2},\frac{i\kappa}{A}+\frac{i\Omega}{A}-\frac{1}{2}\Big).
\end{split}
\end{equation}
Now we take the square modulus of Eq.(\ref{g2}). Note that $\omega$ is the frequency registered by the Minkowski observer in DSR framework and this assumed to be distributed over the continuous range of frequencies $\Omega$ associated with the Fourier transform, such that it peaks at $\Omega=\omega$ \cite{paul}. Hence from now on wards, we take $\omega=\Omega$. By using the definition $\mathcal{P}(\Omega)=\Omega|\mathcal{F}(\Omega)|^2$ \cite{paddy1}, we get the power spectrum for the mode given in Eq.(\ref{g1}) as
\begin{equation}\label{g3}
\begin{split}
 \mathcal{P}_{\kappa}(\Omega)=&\frac{2\pi}{A}\frac{1}{e^{2\pi\Omega/A}+1}\times\\
 &\bigg[\frac{1-\Omega/\kappa-e^{-2\pi\kappa/A}\big(1+\Omega/\kappa\big)}{\big(A^2/4\kappa^2+(1+\Omega/\kappa)^2\big)\big(1-e^{-2\pi\kappa/A}e^{-2\pi\Omega/A}\big)}\bigg].
\end{split}
\end{equation} 
We find that the power spectrum (given in Eq.(\ref{g3})) for the outgoing Dirac modes, detected by the Rindler observer in the DSR framework retains is found to deviate from the well known Fermi-Dirac distribution factor i.e., $(e^{2\pi\Omega/A}+1)^{-1}$, due to $\kappa$ energy scale imposed by DSR. Similar deviation is also seen with the scalar modes in DSR as shown in Eq.(\ref{f7}). 

\begin{figure}[!htb]\centering
\includegraphics[width=0.50\textwidth]{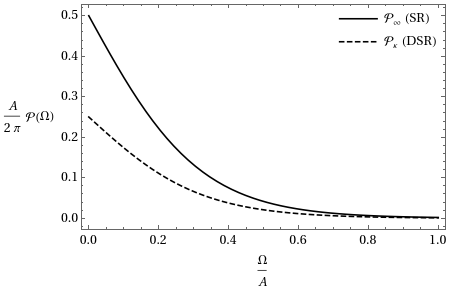}
\caption{Plot of the power spectrum for the outgoing Dirac mode in DSR detected by a Rindler observer} 
\label{fig:unrh1}
\end{figure}

Fig.(\ref{fig:unrh1}) shows the variation of $\frac{A}{2\pi}\mathcal{P}_{\kappa}(\Omega)$ against $\frac{\Omega}{A}$ for the Dirac mode given in Eq.(\ref{g1}). The dotted and solid lines represent the DSR (where we set $\frac{A}{\kappa}\sim 10^{-3}$) and special relativistic (where $\kappa\to\infty$) frameworks, respectively. In this case also we observe a noticeable deviation of the power spectrum (given in Eq.(\ref{g3})) from the standard Fermi-Dirac distribution, especially in the small frequency limit. Similarly in the asymptotic limit, the DSR power spectrum is seen to coincide with the standard ones, as in the previous case of scalar modes. Here also we find that the $\kappa$-energy scale (of the DSR) reduces the value of the usual power spectrum in the special relativistic scenario. 

From Fig(\ref{fig:unrh1}) we find that the deviation of the modified power spectrum, i.e., Eq.(\ref{g3}) from the standard Fermi-Dirac distribution becomes smaller in the larger frequency limit and hence, as in the previous case for scalar modes, here also we assume $\mathcal{P}_{\kappa}(\Omega)$ to have a thermal behaviour in the large frequency limit whose corresponding temperature can be read-off using Eq.(\ref{f8}). Thus the temperature of the thermal bath for the power spectrum obtained in Eq.(\ref{g3}) can be calculated by substituting Eq.(\ref{g3}) in Eq.(\ref{f8}) and we get this temperature as
\begin{equation}\label{g4}
\begin{split}
 T_{\kappa}=&\frac{A}{2\pi}+\\
 &\frac{A^2}{4\pi^2\Omega}\ln\Bigg[1+\frac{\bigg(\Big(\big(1+\frac{\Omega}{\kappa}\big)^2+\frac{A^2}{4\kappa^2}\Big)e^{-\frac{2\pi\Omega}{A}}-\frac{\Omega}{\kappa}\bigg)e^{-\frac{2\pi\kappa}{A}}}{\big(\frac{A^2}{4\kappa^2}+(1+\frac{\Omega}{\kappa})^2\big)\big(1-e^{-\frac{2\pi\kappa}{A}}e^{-\frac{2\pi\Omega}{A}}\big)}\\
 &-\frac{\Big(\frac{3\Omega}{\kappa}+\frac{\big(A^2+4\Omega^2\big)}{4\kappa^2}\Big)}{\big(\frac{A^2}{4\kappa^2}+(1+\frac{\Omega}{\kappa})^2\big)\big(1-e^{-\frac{2\pi\kappa}{A}}e^{-\frac{2\pi\Omega}{A}}\big)}\Bigg].
\end{split}
\end{equation} 
The $T_{\kappa}$ calculated in Eq.(\ref{g4}) is the modified Unruh temperature associated with the Fermi-Dirac distribution in the DSR framework. Here also this temperature is modified due to $\kappa$-dependent energy scale of DSR and this correction terms contain $\Omega$ as in Eq.(\ref{f9}). As in the previous case for the scalar modes, here also the correction terms depend non-linearly on the constant proper acceleration. 

Even though the expression for the Unruh temperature of the thermal bath is same for both the Bose-Einstein and Fermi-Dirac distributions in the standard case, the situation is different when we incorporate the DSR effects. The $\kappa$-dependent corrections to the modified Unruh temperature obtained in Eq.(\ref{f9}) and Eq.(\ref{g4}) are different due to the difference in the correction terms of the corresponding power spectra. 

It is worth noting that the modified Unruh temperature calculated using the power spectrum method and that obtained from the Unruh-De Witt detector method could be different due to the ambiguity in the ordering prescription associated with the non-commutative space-times. At the same time all these expressions for the modified Unruh temperature calculated using various approaches will reduce to the standard expression for the Unruh temperature in the commutative limit. 

\section{Conclusion}

We have studied the Doppler effect within the context of DSR having the $\kappa$-Minkowski NC phase space algebra compatible with the $\kappa$-Poincare algebra in MS basis. We have calculated the expression for the DSR Doppler shifted frequency using the $\kappa$-Lorentz transformations of the momenta coordinate in the DSR and then obtained time-dependent Doppler frequency in DSR using the notion of uniform acceleration. We also calculated the exact expression alternatively, by using the Lorentz transformation and the inverse Darboux mapping, showing that our results are consistent within the underlying DSR framework. 

We have further used the expression for the DSR Doppler shifted frequency to analyse the DSR red-shift. We have shown that the $\kappa$ dependent corrections of this DSR red-shift depends on the frequency of the emitted photon as well, apart from the dependency on the relative velocity (which is present in the special relativistic case also). Another important observation that comes in this analysis is that the non-commutative effects associated with the DSR makes the red shift smaller. Similar results have also been obtained while studying the non-commutative effects on the gravitational red shift \cite{ncred1,ncred2}. Therefore the non-commutative parameters are expected to suppress the red shift parameters. One requires a detailed study and comparisons with the observational data to provide a more generalised conclusion.

We have studied Unruh effect in the DSR by generalising the Doppler shift method to DSR and calculated the power spectrum for the outgoing modes, having a DSR Doppler shifted frequency, detected by a uniformly accelerating observer. This power spectrum in the DSR scenario has been calculated for both the scalar and Dirac particles, separately. In both these cases we find that the effect of the $\kappa$-energy scale, associated with the DSR, is to decrease the value of the standard power spectrum, registered by the Rindler observer. Such a decrease in the response function due to the $\kappa$-energy scale can also be observed in \cite{kim}, where the Unruh effect in $\kappa$-Minkowski space-time has been studied by calculating the response function of a scalar field coupled to a monopole detector moving along a uniformly accelerating trajectory in the $\kappa$-Minkowski space-time.

The power spectrum we obtained while studying the Unruh effect (for both scalar and Dirac particles) in DSR is found to deviate from the standard Bose-Einstein and Fermi-Dirac distributions, due to the presence of $\kappa$ energy scale of the DSR. Such deviations from the standard thermal distribution functions has also been shown perturbatively using the detector method in \cite{kim,ravi1,ravi2}. We further show that this deviations results in the modification of the Unruh temperature, which then leads to non-linear dependency on the proper acceleration (unlike the linear dependence of Unruh temperature on the proper acceleration in the special relativistic scenario). Similar non-linear dependence on the proper acceleration has been shown to arise while studying the Unruh effect for the non-local field theory models defined by the DSR \cite{gim}. Thus we can infer that the study of Unruh effect for the Lorentz violating models (such as DSR, theories with minimal length scales, etc.,) will lead to the deviation of the thermal power spectrum from the standard BE and FD distributions which will result in the modification of Unruh temperature.


\section*{Acknowledgment}

The author would like to acknowledge Prof. Satish Khasa, Director (Research), Dept. of Physics, DCRUST and Prof. Rajendrakumar Anayath for the useful support.

\section*{Data availability}

Data sharing is not applicable to this article as no datasets were generated or analysed during the current study.


\section*{}

\end{document}